\documentstyle[tighten,preprint,aps,epsfig]{revtex}

\begin{document}

\title{Quark model study of the triton bound state}

\author{B. Juli\'a-D\'{\i} az $^{(1)}$, F. Fern\'andez $^{(1)}$,
A. Valcarce $^{(1)}$, and J. Haidenbauer$^{(2)}$}
\address{$(1)$ Grupo de F\' \i sica Nuclear,
Universidad de Salamanca, E-37008 Salamanca, Spain}
\address{$(2)$ Institut f\"{u}r Kernphysik (Theorie), Forschungszentrum J\"{u}lich,
D-52425 J\"{u}lich, Germany.}
\maketitle

\begin{abstract}
The three-nucleon bound state problem is studied
employing nucleon-nucleon potentials derived
from a basic quark-quark interaction. We analyze
the effects of the nonlocalities generated by the quark model.
The calculated triton binding energies indicate that 
quark-model nonlocalities can yield
additional binding in the order of few hundred keV. 
\end{abstract}
\vspace{2cm}

During the last decade the development of quark-model 
based interactions for the hadronic force
has led to nucleon-nucleon ($NN$) potentials that provide 
a fairly reliable description of the on-shell data. 
As a consequence of the internal structure of the nucleon, 
such interaction models are characterized by the presence
of nonlocalities. 
These nonlocalities are
reflected in the off-shell properties  
and emerge from the underlying dynamics in a natural 
way. 

The relevance and/or necessity of considering the nonlocal
parts of nucleon-nucleon potentials in realistic interactions
is still under debate. Indeed, over the past few years
several studies have appeared in the literature which stress 
the potential importance of nonlocal effects for the quantitative
understanding of few-body observables and, specifically, for the triton 
binding energy \cite{Gibson,HH,MSS,Elster,TA92}. However,
the majority of these investigations \cite{Gibson,HH,MSS,Elster}
explore only nonlocalities arising from the meson-exchange 
picture of the $NN$ interaction. The effects of nonlocalities 
resulting from the quark substructure of the nucleon have only
been addressed once so far \cite{TA92} and more systematic
studies are lacking altogether. 

In this work we study the triton binding energy
by means of a local and a nonlocal potential,
derived from the same constituent quark-model. We 
will pay special attention to the nonlocal effects 
originating from the quark model.
The nonlocal potential is derived by means of the resonating group
method (RGM) and the local one is obtained through the 
Born-Oppenheimer approximation. 
These interactions are employed in 
Faddeev calculations of the three-nucleon binding energy. 

\section{Quark-model based $NN$ potentials}
The underlying idea of the quark model we use
is that the constituent quark mass is a consequence
of the spontaneous chiral symmetry breaking (SCSB).
Then, between the chiral symmetry breaking scale ($\Lambda_{CSB}\sim$ 1 GeV) 
and the confinement scale, 
($\Lambda_{C}\sim$ 0.2 GeV) QCD may be simulated 
in terms of an effective theory of constituent quarks
interacting through the Goldstone modes associated 
with the SCSB. Perturbative features of QCD are 
incorporated through the one-gluon exchange
potential. A more extensive description of the 
quark-model Hamiltonian can be found in the literature
\cite{JPG,AVA95,DEN1}.

Based on the quark model Hamiltonian,
 two different procedures have been used in
the literature to obtain baryonic interactions.
The first one is the RGM. It allows to treat the inter-cluster 
dynamics in an exact way once the Hilbert space has been fixed. 
The potential derived from this method contains all
the nonlocalities associated with quark antisymmetry. For the
present study we will make use of the $NN$ potential derived 
through a Lippmann-Schwinger formulation of the 
RGM equations in momentum space \cite{DEN1}. It reproduces the 
$NN$ phase shifts up to 250 MeV lab energy with
quite a good accuracy.

The second method is based on the Born-Oppenheimer approximation. 
It provides a clear-cut 
prescription for removing the nonlocalities while preserving
the general properties of the interaction for lower partial 
waves, i.e., those coming form quark antisymmetry.
This local interaction
has been widely applied to
a great variety of physical problems, obtaining reasonable results.
In general, the phase shifts are reproduced with a comparable accuracy
to the RGM results \cite{RDM1}.

In both cases, for a correct description of the $^1S_0$ phase shift
it is necessary to take into account the coupling 
to the $^5D_0 \ {N\Delta}$ channel \cite{AVA95}, providing 
the required additional attraction. 
In order to achieve almost phase shift equivalence between 
the local and nonlocal interaction models, which is mandatory
if one wants to reliably judge the influence of the nonlocalities,
we have done a fine tuning of the potential parameters. 

The results obtained for the two-body system with the local and nonlocal
potentials are presented in Table \ref{table2} and Figs. \ref{fig1} and
\ref{fig2}. 
The $^1S_0$ and the $^3S_1-{^3D_1}$ phase shifts and the low-energy
scattering parameters as well as the deuteron
binding energy are practically the same for both potential models, and
also in very good agreement with experimental data. 
 
\section{Triton binding energy: Results and discussion}

The three-body system is solved performing a five channel 
Fadeev calculation including the $^1S_0$ and $^3S_1-{^3D_1}$ $NN$
partial waves as input. 
Note that since in our model there 
is a coupling to the $N\Delta$ system,  
a fully consistent calculation would require the inclusion of 
two more three-body channels. However, their
contribution to the $3N$ binding energy is known to be rather small \cite{Hajd}
and therefore we neglect them for 
simplicity reasons. 

To solve the three-body Faddeev equations in momentum space
we first perform a separable finite-rank expansion of the
$NN(-N\Delta)$ sector utilizing the EST method \cite{EST,WSH00,NEMO}. 
In Ref. \cite{NEMO} it was shown that 
with a 
separable expansion of sufficiently high rank, reliable and
accurate results on the three-body level can be achieved. 
In the present case it turned out that separable 
representations of rank 6-8 for $^1S_0-(^5D_0)$ and rank 6 for
$^3S_1-{^3D_1}$, are sufficient to get converged 
results. 
The results of the triton bound state calculation 
are summarized in Table \ref{table4}. 

\begin{figure}[tbp]
\epsfig{file=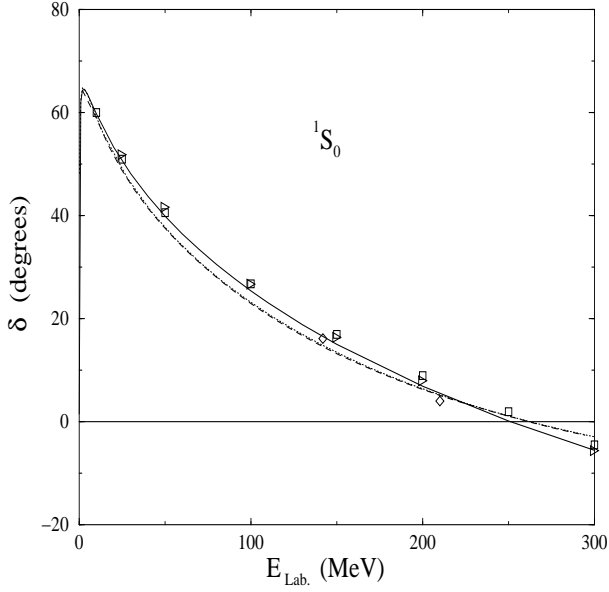, height=8cm,width=8cm}
\caption{$NN$ phase shifts. The solid line stands for the nonlocal
potential, the dashed line corresponds to the local one. The squares,
diamonds and triangles are the experimental data taken from
Refs. \protect\cite{ex1}. The dotted line shows the result of the EST separable
representation of the local model.}
\label{fig1}
\end{figure}

Let us first emphasize that the predicted binding energies
for both models are comparable to those obtained from 
conventional potentials of the $NN$ interaction such as the 
Paris, Bonn, or Nijmegen models \cite{WSH00,xxx}. 
Comparing the values for our local and nonlocal models,
one observes that there is about 150 keV more binding for
the nonlocal potential. Is this the enhancement we can
expect from the nonlocalities due to the quark
substructure of the nucleon? In order to answer this
question we need to go back again to the $NN$ results
and scrutinize the on-shell properties carefully.
For the $^1S_0$ partial wave the differences in the
low-energy scattering parameters and in the phase shift are indeed 
very small, see Table~\ref{table2} and Fig.~\ref{fig1} 
respectively.

\begin{figure}[tbp]
\epsfig{file=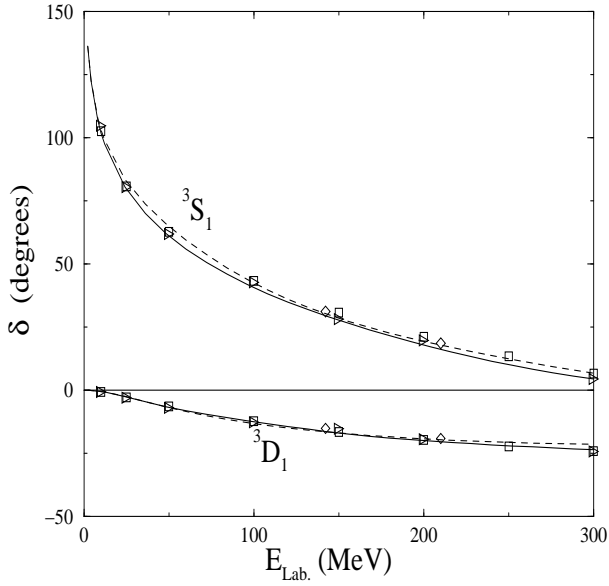, height=8cm,width=8cm}
\caption{$NN$ phase shifts. The solid line stands for the nonlocal
potential, the dashed line corresponds to the local one. The squares,
diamonds and triangles are the experimental data taken from
Refs. \protect\cite{ex1}.}

\label{fig2}
\end{figure}

Unfortunately, for the $^3S_1-{^3D_1}$ partial wave the situation
is much more complicated. While the deuteron binding energy and
also the $^3S_1$ and $^3D_1$ phase shifts are in excellent agreement,
(see Fig.~\ref{fig2}) 
this cannot be said about the mixing 
parameter $\varepsilon_1$.
In this case, 
it is difficult to estimate reliably the effect from the obvious 
deviation from phase equivalence on the triton binding energy.

However, one can clearly separate 
the effects from the two involved partial 
waves. For this purpose, we carried out additional $3N$ calculations where we 
combined the $^1S_0$ of the local model with the 
$^3S_1-{^3D_1}$ of the nonlocal model and vice versa. 
Corresponding results are compiled in Table~\ref{table5} 
where we show triton binding energies (in MeV) for different 
combinations of the local and nonlocal models. 
They strongly suggest that the nonlocalities present in the
$^1S_0$ alone are already responsible for the enhancement of
around 150 keV in the triton binding energy. The shift in the
binding energy 
is independent of whether we use the local or nonlocal version
model for the $^3S_1-{^3D_1}$ partial wave. 
On the other hand, the nonlocalities present in the 
$^3S_1-{^3D_1}$ partial wave seem to even decrease the
binding energy. However, we suspect that here the effect
of the nonlocalities is obscured by the fact that the
two models are not strictly phase equivalent.

In summary, we have calculated 
the three-nucleon bound state problem utilizing $NN$ 
potentials derived from a basic quark-quark interaction. 
One of these potentials was generated by means of the
resonating group method so that nonlocalities resulting
from the internal structure of the nucleon were preserved.
The other potential is based on the Born-Oppenheimer
approximation and is strictly local. These potentials are
made nearly phase equivalent by fine tuning of
some of the model parameters. 
The corresponding calculations of the triton bound state indicate 
that the nonlocalities resulting from the quark substructure of 
the nucleon yield additional attraction and, specifically,
can lead to an increase of the binding energy by up to 200 keV.  
Thus, the effect of those nonlocalities on the three-nucleon
binding energy is certainly appreciable. In particular, 
it is of the same magnitude as the one resulting from 
nonlocalities that occur in the meson-exchange picture of the 
$NN$ interaction. 

\begin{table}[h]
\caption{$NN$ properties}
\label{table2}
\begin{tabular}{llll|llll}
\hline
\multicolumn{4}{c}{ $NN$ Low-energy scattering parameters} &
\multicolumn{4}{c}{ Deuteron properties} \\
\hline
 & &Local  & Nonlocal & & & Local  & Nonlocal  \\
\hline
 $^1S_0$ & $a_s$ (fm) & -23.758 & -23.759 &
\multicolumn{2}{l}{$\epsilon_d$ (MeV)} & -2.2245 & -2.2242   \\
         & $r_s$ (fm) &  2.694 & 2.682 &
\multicolumn{2}{l}{$P_D$ ($\%$)}& 4.79  & 4.85     \\
 $^3S_1$ & $a_t$ (fm) & 5.464  &  5.461& 
\multicolumn{2}{l}{$Q_d$ (fm$^2$)}& 0.280  & 0.276 \\
         & $r_t$ (fm) & 1.779  &  1.820& 
\multicolumn{2}{l}{$A_S$ (fm$^{-1/2}$)}& 0.900  & 0.891 \\
         &            &        &       &
\multicolumn{2}{l}{$A_D / A_S $}&  0.0243 & 0.0257    \\
\hline
\end{tabular}
\end{table}      
 
\begin{table}[h]
\caption{Properties of the three-nucleon bound state.}
\label{table4}
\begin{tabular}{c|ccccc}
&$E_B$ (MeV) &$P_S$ ($\%$) & $P_{S'}$ ($\%$) &$P_P$ ($\%$) &$P_D$ ($\%$)\\
\hline
Local  &  -7.572 &91.413 &1.597 &0.044 & 6.946 \\
Nonlocal& -7.715 &91.493 &1.430 &0.044 & 7.033  \\
\end{tabular}
\end{table}  

\begin{table}[h]
\caption{}
\begin{tabular}{llll}
& & \multicolumn{2}{c}{ $^3S_1-{^3D_1}$} \\
& &Local  & Nonlocal   \\
\hline
 $^1S_0$ &Local & -7.572 & -7.544  \\
         &Nonlocal & -7.745 & -7.715  \\
\end{tabular}
\label{table5}
\end{table}         

\end{document}